\documentclass[11pt,a4paper]{scrartcl}

\usepackage{ILD}

\usepackage[bottom]{footmisc}
\usepackage{feynmf}
\usepackage{multirow}
\usepackage{textpos}

\usepackage{hyperref}
\usepackage{url}
\usepackage{mathtools}
\usepackage{wrapfig}
\usepackage{footnote}
\usepackage{subcaption}
\usepackage{caption}
\usepackage{verbatim}
\usepackage{comment}
\usepackage{indentfirst}
\usepackage{wasysym}
\usepackage{multirow}
\usepackage{xspace}
\usepackage{acronym}

\newcommand{\epem}{e$^+$e$^-$\xspace}
\newcommand{\ggtohad}{$\gamma \gamma \to$ hadrons\xspace}
\newcommand{\qqbar}{q$\bar{\mathrm{q}}$\xspace}

\newlength{\figwidth}
\newlength{\figskip}
\usepackage[math]{cellspace}
\title{Searching for displaced vertices with a gaseous tracker for a future \epem Higgs factory}

\ildphys{2024}{004}

\date{Revision for JHEP\\ \formatdate{10}{01}{2025}}

\addauthor{Jan Klamka}{\institute{}}
\addauthor{Aleksander~Filip~\.Zarnecki}{\institute{}}

\addinstitute{}{Faculty of Physics, University of Warsaw,\\
  Pasteura 5, 02-093 Warsaw, Poland}

\abstract{
This paper presents results of the first full simulation study addressing prospects for observation of long-lived particles (LLPs) with the International Large Detector (ILD), operating at the International Linear Collider (ILC) at $\sqrt{s}=250$\,GeV. Neutral LLP production, resulting in a displaced vertex signature inside the ILD's time projection chamber (TPC), is considered. We focus on scenarios interesting from the experimental perspective and perform a search based on displaced vertex finding inside the TPC volume. Two experimentally challenging scenarios are explored: the first involving very soft final states due to a small mass splitting between a heavy LLP and the dark matter particle to which it decays,
and the second with production of a light and therefore highly boosted LLP
resulting in almost colinear vertex tracks. The expected limits on the signal production cross section are presented for a wide range of the LLP proper lifetimes corresponding to $c\tau$ from 0.1\,mm to 10\,km.
}

\titlecomment{This work was carried out in the framework of the ILD Concept Group} 

\graphicspath{ {./logos/}{./figures/} }

\begin{document}

\titlepage

\section{Introduction}

The Standard Model (SM) of particle physics~\cite{Glashow:1961tr,Weinberg:1967tq,Salam:1959zz,Englert:1964et,Higgs:1964pj,Guralnik:1964eu} provides a comprehensive framework for understanding the interactions among constituents of the Universe discovered so far. 
Despite its remarkable success~\cite{UA1:1983crd,UA2:1983tsx,UA1:1983mne,UA2:1983mlz,ATLAS:2012yve,CMS:2012qbp}, there are a number of theoretical and experimental indications which prompt searches for new physics phenomena Beyond the SM (BSM), such as Dark Matter (DM) particles, additional sources of CP violation, or the origin of neutrino masses. 
An interesting case explored in many recent studies involves new Long-Lived Particles (LLPs), i.e. BSM particles characterized by their macroscopic lifetimes (of the order of picoseconds or higher). One should realize that many such states  occur within the SM (neutrons, muons, kaons, pions, etc.) so there is no explicit reason to expect that all hypothetical BSM particles should decay promptly. States with macroscopic lifetimes naturally appear in many BSM models~\cite{Barbier:2004ez,Strassler:2006im,Kaplan:2009ag,Izaguirre:2015zva,Bauer:2017ris}, and if they exist, could evade many of the standard new physics searches conducted e.g. at the Large Hadron Collider (LHC).
A dedicated event reconstruction and selection approach is needed to search for LLP production, targeting unique signatures such as displaced vertices (with tracks or jets in the final state), disappearing or kinked tracks, strongly ionizing charged particles, non-pointing photons, and more~\cite{Alimena:2019zri}.

Although LLPs have already been considered in a variety of BSM searches at the LHC, no indication for LLP production has been observed so far. The ATLAS and CMS experiments have probed large parts of the parameter space, imposing strong constraints on very massive LLPs for a wide range of lifetimes. Searches for decays inside the muon systems probe decay lengths above 500\,m for LLP masses of 40\,GeV~\cite{CMS:2024bvl,ATLAS:2018tup}, while considering delayed decays constrains the proper decay lengths even at the order of $10^{12}$\,m~\cite{CMS:2017kku}.  Huge masses of up to 2.5\,TeV have also been probed in searches for displaced jets~\cite{CMS:2020iwv} and up to 1.8\,TeV using a displaced vertex with missing momentum signature~\cite{CMS:2024trg,ATLAS:2017tny}.  Lighter LLPs have recently started to gain more interest. In particular, models with LLP masses as small as 400\,MeV have been constrained at the LHC in Higgs boson decays~\cite{CMS:2024bvl,ATLAS:2023cjw} and a recent search for displaced muon pairs probes heavy squark decays to long-lived neutralinos, where the mass splitting between those two is only 25\,GeV~\cite{CMS:2024qxz}. However, in this regime experiments at the LHC suffer from a huge background, which can affect searches for both light and heavier states, in particular in the case of models with compressed spectra (small mass differences between particles in the decay chain).  In order to gain sensitivity, light LLPs must originate from decays of much heavier particles, or the analysis must rely on some model-specific assumptions. 

Small couplings and compressed spectra are exactly the features leading to predictions of LLPs by theoretical models~\cite{Lee:2018pag}, so there is a strong motivation for testing this region beyond the current LHC reach. The recent growing interest in models with so-called feebly interacting massive particles (FIMPs), which often exhibit macroscopic lifetimes or appear alongside LLPs, also points in this direction~\cite{Antel:2023hkf}.  This could be addressed by the future Higgs factories, which can complement searches for LLPs at the LHC. 

According to the 2020 Update of the European Strategy for Particle Physics, an \epem Higgs factory is ``the highest-priority next collider"~\cite{EuropeanStrategyGroup:2020pow}. Currently, there are several proposals for such a machine, the most studied of which are: the International Linear Collider (ILC)~\cite{Bambade:2019fyw}, the Compact Linear Collider (CLIC)~\cite{Aicheler:2012bya}, the Future Circular Collider (FCC-ee)~\cite{FCC:2018evy}, and the Circular Electron Positron Collider (CEPC)~\cite{CEPCStudyGroup:2018rmc}. In this paper, we consider the ILC as a ``reference collider", since it remains the most mature option, with its Technical Design Report published in 2013~\cite{ILC:2013jhg}. The ILC baseline running scenario (H-20)~\cite{Barklow:2015tja} assumes staged operation at two center-of-mass energies (250 and 500\,GeV), with the electron and positron beams polarized longitudinally and luminosity sharing between the four beam helicity configurations. A total of 2\,ab$^{-1}$ and 4\,ab$^{-1}$ integrated luminosity is planned to be collected at the 250 and 500\,GeV stages, respectively. Operation of the ILC at the Z-pole (ILC-GigaZ) and an upgrade to $\sqrt{s}=1$\,TeV are also possible~\cite{Bambade:2019fyw}. In this paper we consider ILC running at 250\,GeV and assume a total integrated luminosity of 2\,ab$^{-1}$.

Although their primary goals are precision measurements of the Higgs boson, top quark, and the electroweak sector, future \epem colliders, with a clean experimental environment and triggerless operation\footnote{The latter is assumed for linear machines; circular colliders, due to much higher collision rate, may require some form of a trigger system.} are also very well suited to look for uncommon signatures and exotic phenomena. 
Prospects for LLP searches at future Higgs factories are one of the focus topics of the ongoing ECFA study~\cite{deBlas:2024bmz}.

This paper focuses on scenarios including light particles and soft final states not (fully) tested at the LHC. Since most LLP signatures involve uncommon track topologies (one or multiple tracks with high displacement or large impact parameter), we consider just two tracks originating from a displaced vertex as a generic case. Very promising in this context is the International Large Detector (ILD) concept~\cite{ILDConceptGroup:2020sfq}, introduced in Section~\ref{sec:ild}. An inclusive analysis is performed, with the selection not optimized for any particular BSM scenario or a specific signature. This is discussed in more detail in Section~\ref{sec:benchmarks}, together with the experiment-orientated approach taken in this study and the selected benchmark scenarios. Section~\ref{sec:overlay} describes beam-induced backgrounds and measures proposed for their reduction, which are followed by a selection aimed to reduce backgrounds from other SM processes, introduced in Section~\ref{sec:sm_bg}. Section~\ref{sec:vtx_finding} presents the performance of a vertex finding procedure developed for this analysis, resulting in the limits on signal production cross section shown in Section~\ref{sec:limits}. The study is summarized in Section~\ref{sec:summary}.

\section{The International Large Detector \label{sec:ild}}

The ILD was originally proposed as one of the experiments for the ILC, although studies are currently ongoing regarding its possible operation also at a circular machine~\cite{Einhaus:2023npl}. 
The ILD tracking system comprises of five subdetectors. The vertex detector (VTX), located closest to the beam pipe, consists of three double layers of silicon pixel sensors. In the central region it is surrounded by two double layers of a silicon internal tracker (SIT), and in the forward region it is extended by seven discs (five of which are double discs) of the forward tracking detectors (FTD). The FTD acceptance starts already at an angle of 4.8 degrees. The main tracker of the ILD is a time projection chamber (TPC) allowing for almost continuous tracking, which can significantly enhance the reach in searches for displaced signatures~\cite{Klamka:2023kmi}. In addition, the TPC is surrounded by one double layer of a strip silicon external tracker (SET). The ILD baseline design is optimized for event reconstruction with the particle-flow approach~\cite{Thomson_2009} based on highly granular calorimeters. For details about the ILD see Ref.~\cite{ILDConceptGroup:2020sfq} and references therein. 

Reconstruction for the ILD concept is performed using iLCSoft~\cite{iLCSoft}, and is based on the Marlin framework~\cite{Gaede:2006pj}. Track reconstruction in the standard reconstruction chain of the ILD~\cite{iLDConfig} is first performed individually in different parts of the detector by dedicated processors~\cite{Gaede_2014}. The \textit{SiliconTracking} processor searches for track segments in the VTX, SIT and FTD, while the \textit{ForwardTracking} processor only considers the FTD. The best subset of tracks reconstructed in the silicon trackers by these two processors is then selected. The \textit{Clupatra} processor is subsequently used for pattern recognition in the TPC. Finally, the \textit{FullLDCTracking} processor attempts to match and combine the reconstructed segments, as well as refitting the obtained tracks, using also hits in the SET. Tracks and segments are fitted at each stage using a Kalman filter.

\section{Analysis strategy and the benchmark scenarios \label{sec:benchmarks}}

\subsection{Strategy}

We study prospects for LLP observation at future colliders from an experimental point of view. Discussed in this paper are two scenarios of LLP production -- a light LLP resulting in boosted, high-$p_T$ SM final state, and a (relatively) heavy LLP, which decays to a heavy DM particle and SM particles.  The benchmark scenarios are not selected based on any particular theoretical model or existing experimental constraints in its parameter space, but rather to provide signatures interesting from the experimental perspective -- potentially challenging or not accessible at the LHC.

We identify three factors that can limit sensitivity to LLP production: soft (but very numerous) beam-induced background, hard SM background, and detector or reconstruction limitations.\footnote{Throughout the paper, \textit{soft} refers to low-$p_T$, and \textit{hard} to high-$p_T$.} Soft signals can be mimicked by the first of these backgrounds and the boosted topologies by the second one, while both scenarios are challenging for detectors and reconstruction techniques. Therefore, with this choice of benchmarks, we test all three potential limitations.

Following this model-independent approach, we do not assume anything about the final state except for the presence of at least one displaced vertex inside the TPC volume, as the most generic case. Hence, the signature considered is at least one displaced vertex in the TPC; we allow other activity in the detector, but ignore it to remain fully model-independent. Only two-prong vertices were considered, as it can be assumed that such a topology will be the most experimentally challenging one. Although a multi-prong topology may bring its own analysis challenges, we expect that finding a displaced vertex with multiple tracks in the final state should not be less efficient than in case of having just two tracks, and the background levels, in particular those coming from random track intersections, should be much smaller.  The vertex-finding procedure is described later in this section, and the selection of vertex candidates in the Sections~\ref{sec:overlay} and \ref{sec:sm_bg} of this paper.

We use particular models to generate samples for the study; however, we do not make any model-specific assumptions in the analysis, in order to make the results relevant for any model providing the considered, or similar, signatures. The selection (described in detail in Sections \ref{sec:overlay} and \ref{sec:sm_bg}) is therefore \textit{not} optimized for a particular model or scenario and uses only variables related to displaced vertices.

\subsection{Benchmark scenarios}

\begin{figure}[bt]
	    \centering
	 	 \begin{subfigure}{0.35\textwidth}
	 	 	\centering
	 	 	\includegraphics[width=\linewidth]{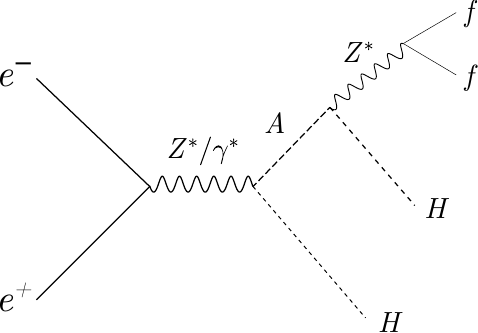}
	 	 \end{subfigure} \qquad
	 	 \begin{subfigure}{0.35\textwidth}
	 	 	\centering
	 	 	\includegraphics[width=\linewidth]{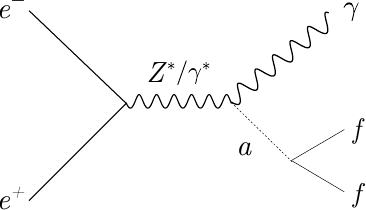}
	 	 \end{subfigure}
	 	 \caption{Leading Feynman diagrams corresponding to the processes used to simulate the considered benchmark scenarios. Left: IDM neutral scalar pair-production with A as the LLP; H is invisible and the Z boson is virtual for small $\Delta m_{AH}$. Right: Associated production of a long-lived ALP, a, with a photon.}
	 	 \label{fig:scenarios}
\end{figure}

The first class of analyzed scenarios involves low-$p_T$ tracks in the final state, whose combined momentum does not extrapolate close to the interaction point (IP).  The model used to generate events with this signature was the Inert Doublet Model (IDM)~\cite{Kalinowski:2018ylg}, which introduces four additional scalars: H$^\pm$, A, and H, which could be lighter or heavier than the SM-like 125\,GeV scalar h. 
The lighter neutral scalar H is stable and is a DM candidate. The neutral scalars can be pair-produced in \epem collisions, with a predominant further decay of A $\to$ Z$^{(*)}$H, as shown in Figure~\ref{fig:scenarios} (left). If the mass splitting $\Delta m_{AH}=m_A - m_H$ between A and H is small, the Z boson is highly virtual and A can be long-lived due to the small decay phase space available. 
Because H is invisible and escapes undetected, only the Z$^{(*)}$ decay products are expected to be measured inside the detector. The small mass splitting and the sizable mass $m_A$ of the A boson implies that the final-state tracks are soft and are not pointing to the IP. We consider four benchmark scenarios with fixed LLP mass $m_A=75$\,GeV and decay length $c\tau=1$\,m, and different $\Delta m_{AH}=1,2,3,5$\,GeV. 

The second type of scenario under consideration was the production of a very light LLP, decaying into strongly boosted and almost colinear tracks. It was modelled using the photon-associated axion-like particle (ALP) production process, as shown in Fig.~\ref{fig:scenarios} (right). 
In many scenarios, ALPs with macroscopic lifetimes for $\mathcal{O}$(GeV) mass scales could be copiously produced in \epem collisions~\cite{Schafer:2022shi}. %
For this signature, we also select four benchmarks, with ALP masses $m_a=0.3,1,3,10$\,GeV and decay lengths $c\tau=10\cdot m_a$\,mm/GeV to maintain large number of decays inside the detector volume.\footnote{Final results of this study will be presented for a wide range of LLP decay lengths. However, by generating signal samples with decay lengths optimal from the detector acceptance point of view we are able to preserve high statistical precision of the result also after sample reweighting to different LLP lifetimes; see Sec.~\ref{sec:limits} for details.}

\subsection{Analysis framework \label{sec:framework}}

For each benchmark scenario, samples of 10,000 events were generated with \textsc{Whizard} 3.0.1 and 3.1.2~\cite{Kilian:2007gr}, for IDM and ALPs respectively, for the ILC running at$\sqrt{s}=250$\,GeV. 
Full simulation of the detector response was based on the ILD-L
detector model~\cite{ILDConceptGroup:2020sfq} implemented in the  \textsc{DD4hep}~\cite{Frank_2014} framework and interfaced with \textsc{Geant4}~\cite{AGOSTINELLI2003250}. Event samples were processed on the grid with the ILCDirac interface~\cite{ilcDiracLC}. 

Reconstruction was performed using iLCSoft~v02-02-03~\cite{iLCSoft} and based on the Marlin framework~\cite{Gaede:2006pj}. The standard ILD reconstruction chain~\cite{iLDConfig} was used, however, selection cuts based on impact parameters $d_0$, $z_0$ (in the XY plane and Z coordinate, respectively~\cite{Kramer:2006zz}) of the reconstructed tracks have not been applied at the final step of track reconstruction, in the \textit{FullLDCTracking} processor~\cite{Gaede_2014}. For this reason, the SM background samples had to be reprocessed, which resulted in a reduced number of analyzed events, compared to most of the ILD studies.

For this study, we designed a dedicated procedure for vertex finding inside the TPC volume. It is meant to be as simple and general as possible. If a track starts inside the TPC, its direction (and hence the charge) is ambiguous, i.e. 
it can be reconstructed as an oppositely charged track traveling in the other direction. For this reason, for each pair of tracks in an event, the algorithm considers four possible hypotheses of the track directions.
The hypothesis chosen is the one for which the tracks have opposite charges and the distance between their first hits is the smallest. Then, the vertex is placed in between the points of the closest approach of track helices, requiring the distance between the helices to be less than 25\,mm. A schematic view of the vertex placement can be seen in Figure~\ref{fig:selection_vars} (Sec.~\ref{sec:overlay}), which illustrates the variables used for the selection.

Only decays of ALPs and Z$^{(*)}$ to muon pairs were generated. This allowed for faster detector response simulation, since muons do not produce showers inside calorimeters as opposed to electrons or pions. 
However, it should be noted that this choice does not make our search less general. The analysis relies exclusively on tracks reconstructed from tracker hits, without considering calorimeter or muon detector signals.
Furthermore, displaced multi-track final states are also included in the analysis (vertices formed by more than two tracks are not rejected), despite the fact that we do not include them explicitly in the signal samples. As a result, we find that both two- and multi-track displaced vertices contribute significantly to the SM background (see Sections~\ref{sec:overlay} and \ref{sec:sm_bg} for details). Since they involve not only muons, but also electrons and pions, it confirms that using muon decays for signal samples does not produce any significant bias.

The analysis is performed in an approach similar to an anomaly search, where the signal can take any form of tracks forming displaced vertices and a discovery would be implied by an excess in the number of reconstructed vertices. As shown in more detail in the following sections of the paper, the backgrounds hardly exhibit any process-dependent behavior, except for a distinction between low- and high-momenta events, and between processes with different numbers of objects expected in the final state. The vast majority of background sources originate from random coincidences of tracks, single neutral, long-lived SM particles, and secondary interactions with the detector material. Therefore, any SM process with a high cross section and a large activity inside the detector is a potential background source.

\section{Beam-induced backgrounds \label{sec:overlay}}

\begin{figure}[bt]
	    \centering
	 	 \begin{subfigure}{0.49\textwidth}
	 	 	\centering
	 	 	\includegraphics[width=\figwidth]{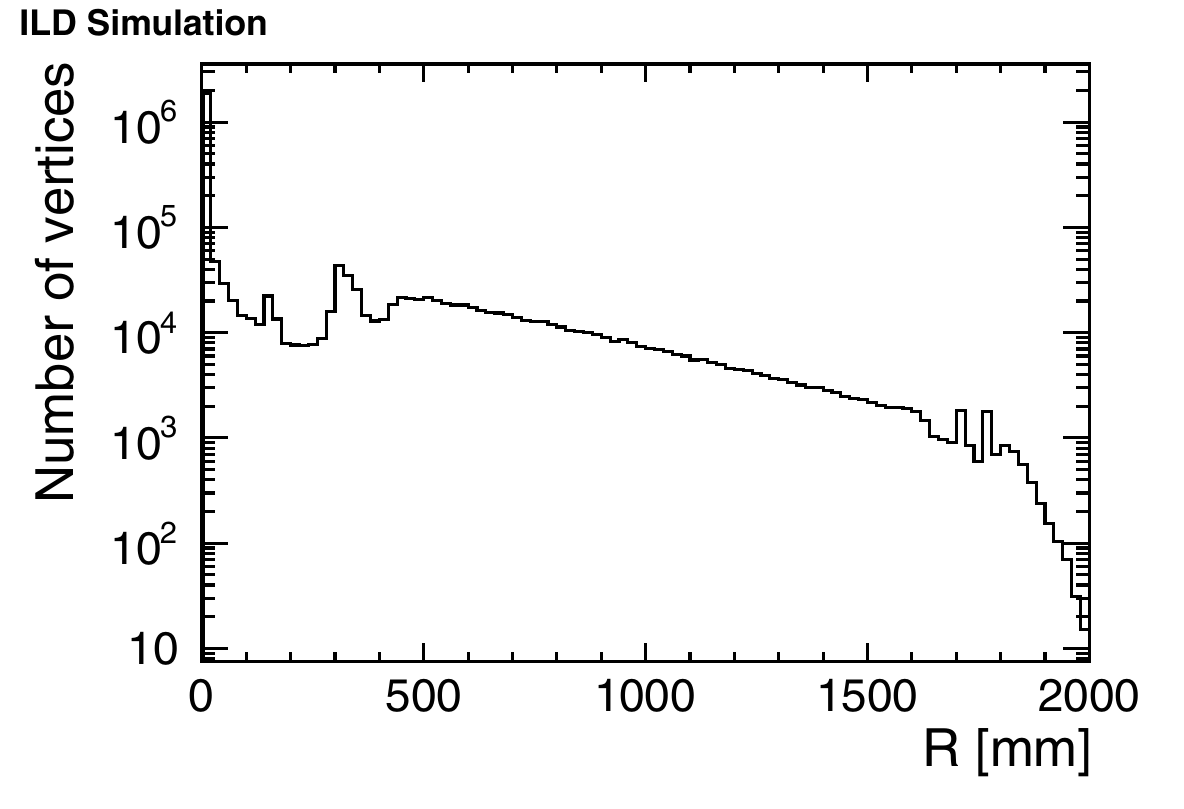}
	 	 \end{subfigure}%
	 	 \begin{subfigure}{0.49\textwidth}
	 	 	\centering
	 	 	\includegraphics[width=\figwidth]{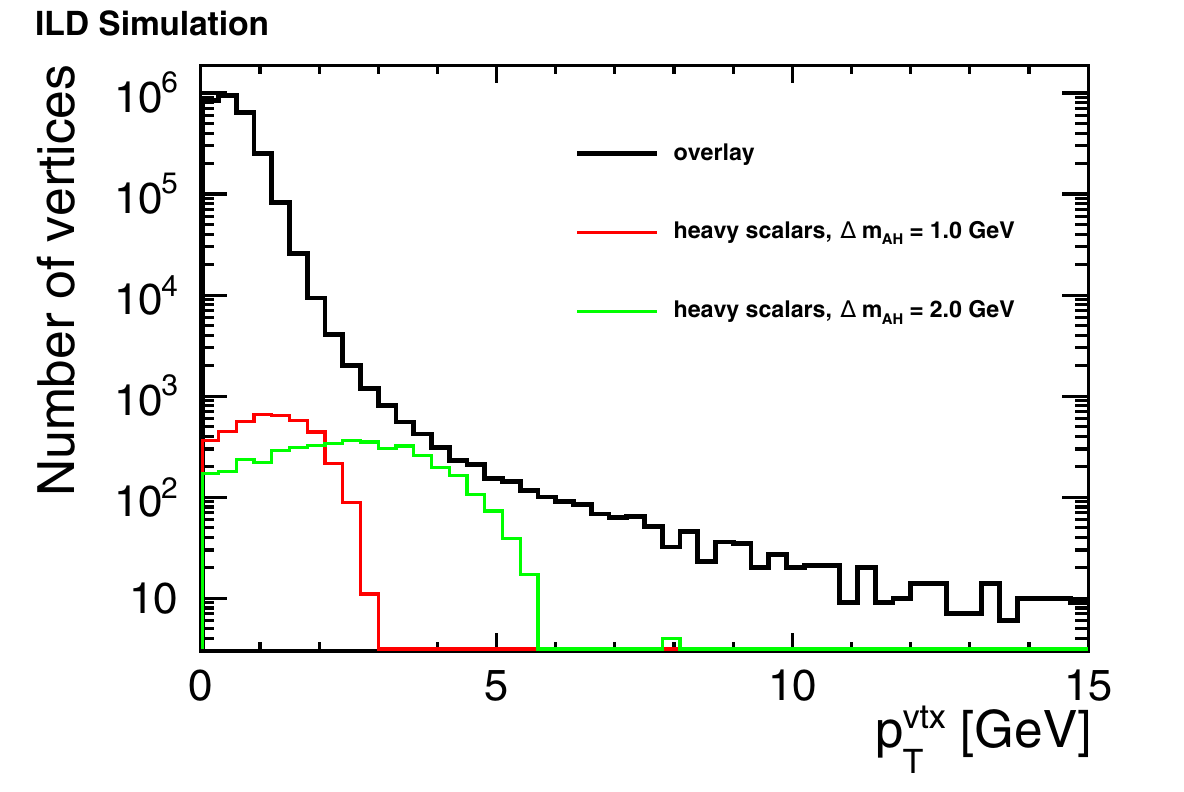}
	 	 \end{subfigure}
  \vspace*{\figskip}
	 	 \caption{Left: Number of displaced vertices found in the overlay sample as a function of distance from the beam axis. Right: Total transverse momentum of tracks coming from a displaced vertex for the overlay (black) and scalar pair-production with $\Delta m_{AH}=1$\,GeV (red) and $\Delta m_{AH}=2$\,GeV (green). All histograms are normalized to the number of simulated MC events and correspond to no selection applied at all, other than requiring a maximum distance between track helices.  }
	 	 \label{fig:background}
\end{figure}

At linear \epem colliders the beams need to be very strongly focused at the IP to reach the required luminosity. Consequently, the charge density at the IP is very high. Therefore, besides quasi-real photons radiated by individual electrons (as described by Weizs\"acker-Williams approximation), bunches are accompanied by synchrotron radiation caused by an interaction with the opposite beam, the so-called beamstrahlung. Both quasi-real  and beamstrahlung photons emitted by one of the bunches can interact with those from the opposite bunch. Hence, each bunch crossing (BX) at a linear collider is a source of low-$p_T$ hadron photoproduction ($\gamma\gamma\to$ hadrons) and incoherent \epem pairs due to beam-induced photon interactions. In the ILD operating at the 250\,GeV ILC, around 1.55 \ggtohad events are expected on average per BX. In addition, $\mathcal{O}(10^5)$ incoherent pairs are produced per BX. However, they are predominantly emitted in the forward direction and only a small fraction of them enters the detector and can be measured. Both of these processes occur in the detector simultaneously with physical hard events (and hence they are typically called \textit{overlay} events). However, with an order of $10^{11}$ BXs expected at the ILC per year,
these events are a potential source of a standalone background. In particular, they are relevant for the signal scenarios considered in this study, which involve soft final states. Figure~\ref{fig:background} (left) presents a distribution of displaced vertices as a function of the distance $R$ from the beam axis. There is an overwhelming number of vertices found in the overlay samples, most of which are fake (e.g. originate from randomly intersecting tracks). Others result from particle interactions in the detector: the first small peak above 100\,mm aligns with the first layer of SIT, while the second one, above 300\,mm, with the last SIT layer and the inner TPC wall. Additional peaks around the outer TPC wall and the SET are visible above 1.7\,m. Figure~\ref{fig:background} (right) shows the distribution of the total transverse momentum, $p_T^{vtx}$, of tracks forming a displaced vertex candidate, comparing overlay events to two of the considered signal scenarios. It is visible that vertices in the overlay sample occupy the same kinematic region. These distributions show that \ggtohad and incoherent \epem pairs events cannot be neglected as a background source in this study, and will require a dedicated mitigation strategy. It should be noted that the numbers of vertices presented in Figure~\ref{fig:background} are not normalized and represent the absolute numbers of vertices in the MC samples. For the background, this corresponds to around $3.6\cdot10^{-6}$ of the total luminosity. For the signal, we do not refer to cross section predictions of any model, to which the distributions could be normalized. The purpose of these plots is only to illustrate the respective kinematic properties of signal and background events. 

Soft hadrons can be produced in photon-photon interactions both from beamstrahlung ($\gamma^B$) and from the Weizs\"acker-Williams spectrum ($\gamma^W$), so four channels are possible. The number of events occurring per single BX for each of them is a random number from the Poisson distribution with a given mean, $\mu_P$, which is used in the standard ILD reconstruction chain when overlaying the background on signal events. Table~\ref{tab:overlay} presents these expected values for the 250\,GeV ILC, together with the number $N_{evt}$ of events used in this study for each channel.  The incoherent \epem pairs are also included in the analysis. To do so, 99,000 BXs were generated with GuineaPig~\cite{Schulte:1999tx}. For each of these BXs, only the particles that would potentially hit any of the tracking detectors were written out and fully simulated. From this pool of BXs, one was picked at random for each physics event, and overlayed before reconstruction. This set of BXs was also used as a standalone background, similar to the hadron photoproduction.

\begin{table}[bt]
    \centering
    \caption{Average number, $\mu_P$, of overlay events expected per single BX for different channels and the number, $N_{evt}$, of corresponding MC events used in the analysis.}
    \label{tab:overlay}
    \begin{tabular}{ccc} \hline
         channel&  $\mu_P$&  $N_{evt}$\\ \hline
          $\gamma^B\gamma^B$&  0.8298&  1,418,000\\ 
          $\gamma^B\gamma^W$&  0.2972&  889,000\\ 
          $\gamma^W\gamma^B$&  0.2975&  911,000\\ 
          $\gamma^W\gamma^W$&  0.1257&  519,000\\ \hline
          tot. \ggtohad&  1.55&  3,837,000\\ \hline
    \end{tabular} 
\end{table}

\subsection{Preliminary cuts \label{sec:pre-cuts}}

\begin{figure}[bt]
	    \centering
	 	 \begin{subfigure}{0.49\textwidth}
	 	 	\centering
	 	 	\includegraphics[width=\figwidth]{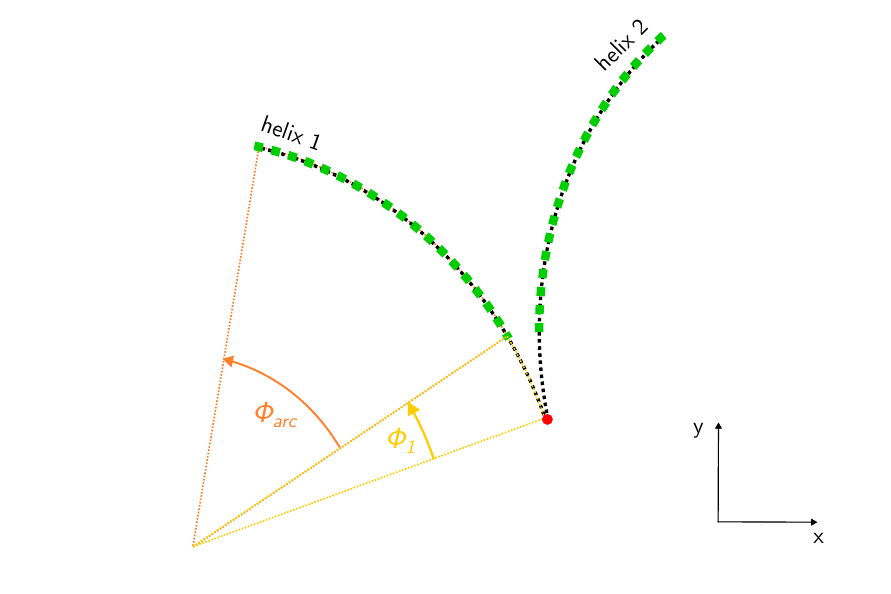}
	 	 \end{subfigure}%
	 	 \begin{subfigure}{0.49\textwidth}
	 	 	\centering
	 	 	\includegraphics[width=\figwidth]{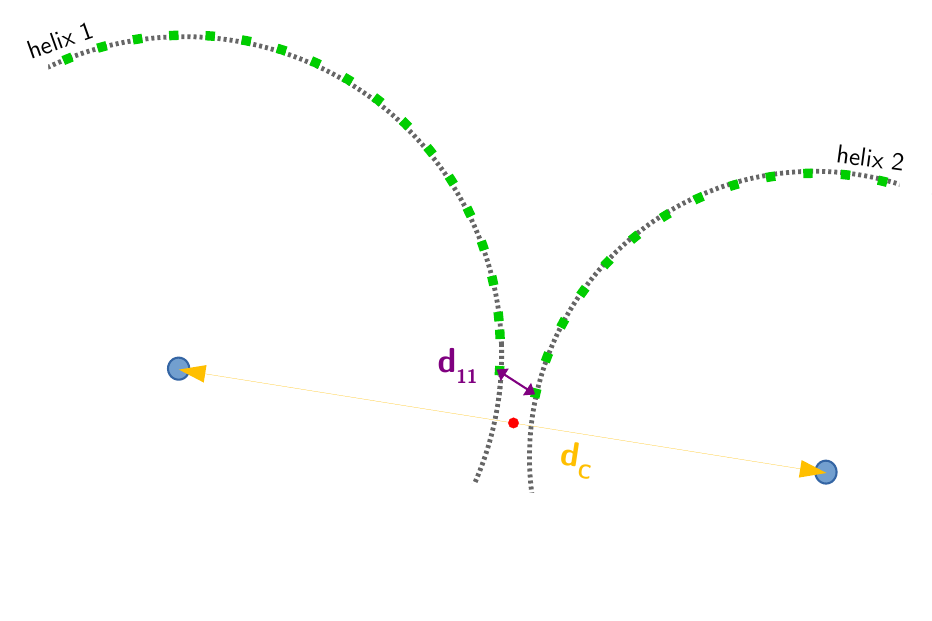}
	 	 \end{subfigure}
  \vspace*{\figskip}
	 	 \caption{Schematic illustration of variables used for the overlay selection. Helices are represented by black dotted lines and detector hits by small green rectangles. Left: Projection of helices onto the XY plane and the angular variables $\phi_{1}$ and $\phi_{arc}$ used in the preselection. Right: Variables $d_C$ and $d_{11}$ that combined are used in the final selection. Small blue circles represent centers of projections of helices onto the XY plane. Note that $d_{11}$ is a three-dimensional distance. The red dot corresponds to a vertex candidate.}
	 	 \label{fig:selection_vars}
\end{figure}

Highly displaced tracks, not pointing to the IP region, are not considered in the standard event reconstruction at the ILD, and therefore a new, dedicated selection procedure had to be implemented. It was designed to be as general as possible, accounting for the background properties, as described in detail below. We apply three sets of preliminary cuts to reduce the number of fake vertices of different origin. 
The first set of ``quality cuts'' is based on variables describing track pair geometry and is aimed at reduction of nonphysical vertices coming from reconstruction inaccuracies or random coincidences. 
\begin{itemize}
    \item Sometimes a charged particle will scatter at unusually large angle at one point in the detector, and the standard track-reconstruction will split the track in two. Track splitting might also occur during reconstruction, when merging of a curling track segments fails. In this analysis, this would be interpreted as a displaced vertex at the point where the track was split. However, the two track segments will be close to colinear at the split-point, and the curvature of the segments will be close to identical. Therefore, we restrict the track pair opening angle $\alpha$ to $\cos{\alpha} > -0.6$ and the ratio of the curvature of the tracks to $\left| \Omega_1 \right| / \left| \Omega_2 \right| < 0.94$, assuming $\left| \Omega_1 \right| < \left| \Omega_2 \right|$ (following definitions of helix parametrization in Ref.~\cite{Kramer:2006zz}). 
\item In some cases, reconstruction of one of the segments of a curling track fails completely, which can lead to formation of poor quality, short tracks, often with unexpectedly large $p_T$\footnote{In the standard reconstruction, these are rejected by cuts on $d_0$ and $z_0$ parameters, which were removed for the purpose of this analysis.}. Such tracks are more likely to pick up a second track in the event to form a vertex. To reject such vertices, it was required that both tracks should have the number of degrees of freedom in the track-fit exceeding 40, a number increased to 70 for the highest $p_T$ track of the vertex, if its $p_T$ was above 1.5 GeV.
\item To further suppress vertices coming from  accidental intersections of uncorrelated tracks, the vertex should be close to or before the first  point measured on the track. To account for tracks with high $p_T$, almost perpendicular to the beam axis, as well as low-$p_T$ tracks curling many times along the Z axis, we use two complementary sets of cuts, depending on the distance $z_{12}=|z_{\mathrm{first}}-z_{\mathrm{last}}|$ between the first and the last track hit in the Z coordinate. 
For tracks with $z_{12}\le 100$\,mm, the ratio of azimuthal angular distance $\phi_{1}$ from the vertex to the first hit, relative to the total angle $\phi_{arc}$ from the first to last hit of the track,  $\phi_{1} / \phi_{arc}$, should not exceed 0.05 (see Figure~\ref{fig:selection_vars}). For tracks with  $z_{12}>100$\,mm, the requirement is set on the difference $z_{1}=z_{\mathrm{first}}-z_{\mathrm{vtx}}$ in Z coordinate between the first hit and the vertex, relative to the $z_{12}$. The value of $\mathrm{sgn}(p_z)\cdot z_{1} / z_{12}$ must be higher than $-0.01$, where $p_z$ is the Z component of the track momentum. Note that $z_{1}$ can be  either positive or negative depending on the track direction, so with a perfect reconstruction of the vertex position, this expression should always be positive.

\end{itemize}

An important source of background is interactions of both charged and neutral particles with the detector material producing (one or more) secondary, displaced tracks emerging from a displaced vertex. As we consider two direction hypotheses for each track, a charged particle producing a single secondary track can also look like a vertex, with one of the tracks going back towards the IP. It is worth noting that this is congruent with a potential signature of \textit{charged} LLPs, commonly explored in many searches as a so-called kinked track.  Interestingly, it seems our vertex-finding procedure is sensitive to such signals as well. However, we here focus on the neutral LLPs and regard these vertices as a part of the background. To remove vertices from secondary interactions, a second set of requirements is imposed.
\begin{itemize}
    \item The search region is restricted to the vertex radii in the range 0.4-1.5\,m, since secondary interactions occur mainly (but not only) with the inner and outer walls of the TPC.
    \item Vertices resulting from the interaction of charged particles with the detector material are first rejected by requiring no additional tracks that end within 30\,mm of the vertex and no tracks with $d_0 < 10$\,mm 
    that pass within 30\,mm of the vertex.
    \item If at least one of the tracks at a vertex 
    has $d_0 < 50$\,mm
    we require that its first hit (the one closest to the reconstructed displaced vertex) has a smaller radius $R_{f}$ than its last hit $R_{l}$. This cut is targeting "kink" signatures: the vertex and secondary track reconstructed as an effect of scattering of charged particles.
\end{itemize}
The neutral particles producing two or more secondary tracks are considered as mostly irreducible background, partly mitigated
by the first cut. That is due to the model-independent approach, in which we do not want to exclude scenarios with LLP decays to more than two charged particles, displaced jet signatures in particular, even though they are not analyzed as such.

Finally, with a third set of preselection cuts, we remove vertices corresponding to decays of V$^0$ particles (a collective term for long-lived neutral hadrons and converting photons).
\begin{itemize}
    \item Matching of vertex candidates with the output of a dedicated ILD processor for V$^0$ identification is performed. The LLP vertex candidate is rejected if there is a vertex reconstructed by the processor closer than 30\,mm.
    \item We exclude vertices with an invariant mass $M_{inv}^{\pi\pi}$ (assuming the tracks are pions) in a window of $\pm 50$\,MeV around the K$^0$ mass and $M_{inv}^{\pi p / p\pi}$ (assuming one track is a pion and the other is a proton and vice versa) around the $\Lambda^0$ mass. 
    \item For the assumption that both tracks are electrons, we reject vertices with a mass $M_{inv}^{ee}$ less than 150\,MeV. A wide window aims to account for poorly reconstructed and almost colinear tracks from highly displaced photon conversions.
\end{itemize}
Cuts on the invariant mass of a pair of tracks for different track mass hypotheses are required for more efficient background suppression as the processor for V$^0$ identification has finite efficiency and prioritizes purity. With the overwhelming number of overlay events, the remaining V$^0$ particles not identified by the algorithm still constitute a substantial background for rare processes.

The preliminary selection is summarized in Table~\ref{tab:cut-flow}, which shows the vertex finding rates in the overlay events, and its efficiencies for selected signal scenarios, after consecutive sets of cuts. For the background, the ratio of the number of events with at least one identified vertex to the number of all events in the MC sample is shown. For the signal, the ratio of the number of correctly identified vertices to all events is presented, where the vertex is considered ``correct'' if it is reconstructed within 30\,mm of the true decay vertex. The first row in Table~\ref{tab:cut-flow} corresponds to the requirement of the maximum 25\,mm distance between track helices, imposed at the beginning of the vertex finding procedure (see Section~\ref{sec:framework}). The following rows correspond to the subsequent sets of cuts described above.

\setlength{\tabcolsep}{4pt}
\renewcommand{\arraystretch}{1.1}
\begin{table}[bt]
\centering
\caption{Summary of vertex-finding rates for the beam-induced backgrounds and finding efficiencies in selected signal scenarios after subsequent sets of preliminary cuts. 
For details, please refer to the description of cuts in Section~\ref{sec:pre-cuts}.}
\begin{tabular}{l|c|cccc}
\hline 
 &  Overlay & \multicolumn{4}{c}{Signal efficiency [\%]} \\
\cline{3-6} 
Cut set & rate [\%] & $\Delta m_{AH}=1$\,GeV & $\Delta m_{AH}=2$\,GeV & $m_a=0.3$\,GeV & $m_a=1$\,GeV \\
\hline 
Dist. between helices           & 37.1    & 72.5 & 77.5  & 28.2 & 74.8  \\
Quality cuts          & 1.7     & 58.3  & 64.6  & 17.0  & 56.7  \\
Secondary interact.     & 0.6     & 52.6 & 57.9 & 14.2  & 48.4 \\
V$^0$ particles       & 0.2     & 41.4 & 55.2 & 7.4  & 48.4 \\
\hline 
\end{tabular}
\label{tab:cut-flow}
\end{table}

\subsection{Final selection}

The preliminary cuts described above allow for background reduction by a factor of around 500, down to the level of 0.21\%. However, due to the huge expected number of approximately $10^{12}$ BXs in the experiment, a suppression factor at the level of at least 10$^{-9}$ or higher is needed to achieve a good sensitivity to rare BSM signals. 
Since it is not possible to generate an event sample of a size corresponding to the expected number of BXs in the experiment, a direct determination of the reduction factor is not possible, and can only be estimated assuming factorization, as a product of individual selection cut efficiencies. However, this will only provide a reliable result if the variables used in the selection are statistically independent. 
Therefore, we select uncorrelated variables for the final selection cuts so that the total reduction factor can be obtained by multiplying all cut efficiencies. 

\begin{figure}[bt]
	    \centering
	 	 \begin{subfigure}{0.49\textwidth}
	 	 	\centering
	 	 	\includegraphics[width=\figwidth]{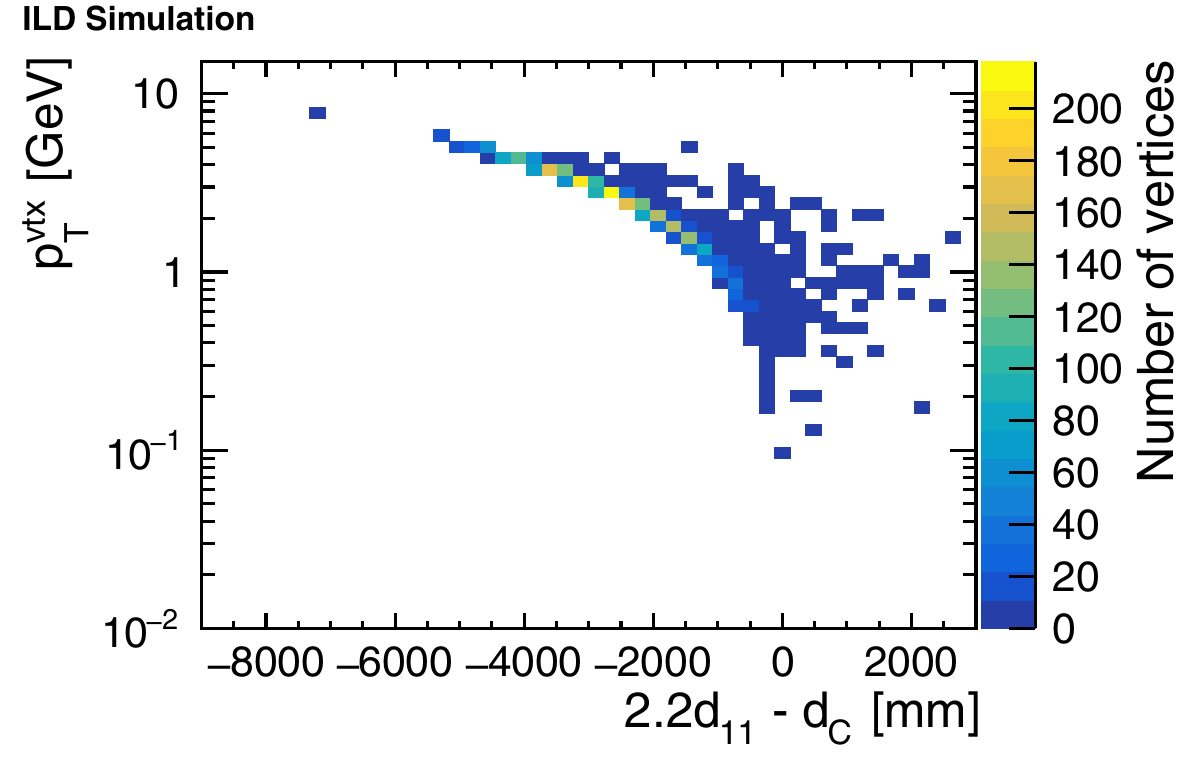}
	 	 \end{subfigure}%
	 	 \begin{subfigure}{0.49\textwidth}
	 	 	\centering
	 	 	\includegraphics[width=\figwidth]{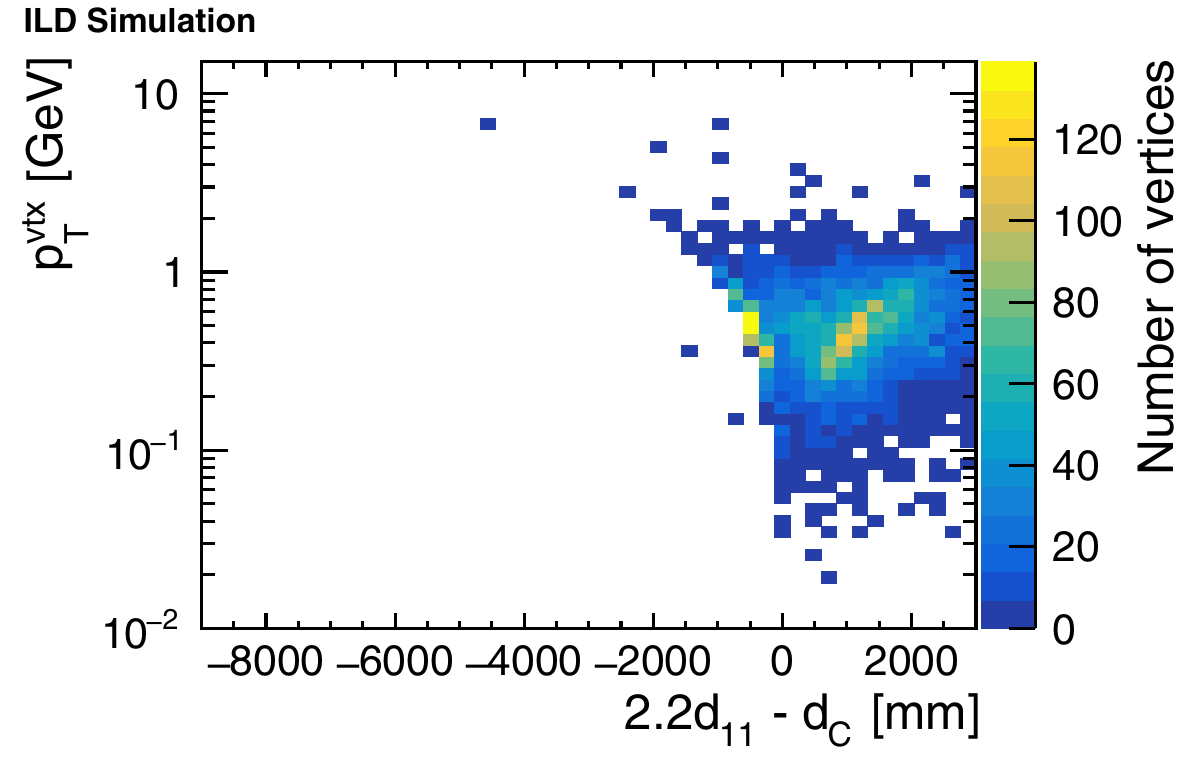}
	 	 \end{subfigure}
	  \vspace*{\figskip}
 	 \caption{Distribution of $p_T^{vtx}$ as a function of $d_h$ variable for a heavy scalars signal scenario with $\Delta m_{AH}=2$\,GeV (left) and all unweighted overlay background events that survive the preliminary cuts (right). For the signal (left) these variables show strong correlation, but for the background (right) only very small correlations are visible for a few events at the lowest values of $d_h$. Both variables are used for the final selection aimed to suppress the background from overlay events; see text for details. }
	 	 \label{fig:dh}
\end{figure}

We use two variables. The first is the total transverse momentum $p_T^{vtx}$ of the pair of tracks. The background is expected to occupy the region of very low $p_T$, as shown in the right plot of Figure~\ref{fig:background}. Tracks in the LLP signals should also come out of a single point, so the distance $d_{11}$ (in three dimensions) between their first hits in the TPC should be small. On the other hand, helix-circles (projections of track helices onto the XY plane) of these tracks should not overlap, so the distance $d_C$ between their centers should be large. A schematic visualization of $d_{11}$ and $d_C$ is shown in Figure~\ref{fig:selection_vars} (right). To reduce the number of selection criteria and avoid correlation between these two variables, we define a new variable $d_h \equiv 2.2d_{11} - d_C$, the distribution of which is shown in Figure~\ref{fig:dh} for the signal (left) and background (right) as a function of $p_T^{vtx}$, after the preliminary selection. The factor 2.2 results in the best signal-background separation, corresponding to the optimal linear discriminant in the $d_{11}$-$d_C$ plane. 
As visible in the distribution on the left, for true vertices the $d_h$ and $p_T^{vtx}$ variables are strongly correlated. However, only little or no correlation should be observed for any fake vertices from random coincidences remaining in the overlay sample, as can be seen in the right plot.
We choose $d_h<-2000$\,mm and $p_T^{vtx}>1.9$\,GeV as the optimal cuts. To mitigate the effect of limited statistics, we estimate the reduction factor for each of these variables from distributions fitted to the respective histograms of $p_T^{vtx}$ and $d_h$. The functions used were, respectively, log-normal distribution for $p_T$ and a sum of Crystal Ball functions, $c_1(x)+c_2(-x)$, for $d_h$. This is performed for all overlay events that survive the preliminary selection, weighted by the corresponding expected values $\mu_P$ shown in Table~\ref{tab:overlay}. Cuts on the $p_T^{vtx}$ and $d_h$ result in selection efficiencies of $1.73\cdot10^{-3}$ and $3.48\cdot10^{-5}$, respectively. As each of these factors is determined on the sample of events after the preliminary cuts, this can be combined with the preliminary selection efficiency, which gives the total reduction factor of $1.26\cdot10^{-10}$.

\section{Backgrounds from physical events \label{sec:sm_bg}}

The SM processes from hard \epem scattering (listed in Table~\ref{tab:cross_sec} together with their cross sections at 250\,GeV ILC) are also taken into account. To increase the production efficiency, the samples were generated with $\pm100\%$ polarization and later reweighted, according to the assumed beam helicity sharing in the ILC H-20 scenario, with $\pm80\%$ electron and $\pm30\%$ positron beam polarization~\cite{Bambade:2019fyw}. 
Preliminary tests have shown that the main sources  of displaced vertices in this case are, similar to overlay events, the interactions with the detector material and decays of long-lived hadrons, which happen predominantly inside the jets. Therefore, one could expect that the 
vertex finding rate for background, defined as a fraction of events with at least one identified vertex to all events, 
does not depend on a physical process itself and neither on the beam polarization, but rather on a number of jets produced. Because rerunning track reconstruction for all SM samples is very time and resource consuming, we obtain the selection efficiencies directly for the most significant processes (those with the highest cross sections),  two-fermion (\epem $\to$ \qqbar) and four-fermion (\epem $\to \mathrm{q\bar{q}q\bar{q}}$) hadronic channels. Then, to estimate the level of background for a complete set of SM processes, we use the \qqbar vertex finding rate for other \epem channels, as they also involve two jets in the final state. 
However, we also take into account the hard scattering of $\gamma^{B/W}\gamma^{B/W}$, which we simulate as well, since this background might in principle contain different sources of (fake) displaced vertices.
Six-fermion and Higgs boson production, as well as e$^\pm \gamma$ interactions, were neglected because of their much smaller cross sections.

\begin{table}[bt]
    \centering
    \caption{Cross sections for the SM processes at $\sqrt{s}=250$\,GeV considered in this study, for different setups of the beam polarization, assuming $\pm100\%$ polarization, and different photon sources for $\gamma\gamma$ processes.}
    \label{tab:cross_sec}
    \begin{tabular}{ccccc} \hline
         sgn(P(e$^-$), P(e$^+$)) &  ($-,+$)&  ($+,-$)&  ($-,-$)& ($+,+$)\\ \hline
         channel& \multicolumn{4}{c}{$\sigma$ [pb]} \\ \hline
         qq&  127.966&  70.417&  0& 0\\ 
          qqqq&  28.66&  0.97&  0& 0\\ 
          qq$\ell\nu$&  29.043&  0.261&  0.191& 0.191\\ 
          qq$\ell\ell$, qq$\nu\nu$& 2.715& 1.817& 1.156& 1.157\\ \hline \hline
          process& BB&  BW&  WB& WW\\ \hline
          $\gamma^{B/W}\gamma^{B/W}$& 42.15& 90.338& 90.12& 71.506\\ \hline

    \end{tabular} 
\end{table}

First, \qqbar is considered as the dominant background, and the selection established for the reduction of overlay events is applied.
To further reduce background from coincidences of random tracks within high $p_T$ hadronic jets,
a separate cut on $d_{11} < 50$\,mm is applied. We will refer to this whole set of cuts as the \textit{standard} selection. It yields the vertex finding rate of $(7.99\pm0.68)\cdot10^{-4}$ for the \qqbar sample. 
The rate obtained for the qqqq sample  was found to be $(1.486\pm0.094)\cdot10^{-3}$, which is a factor of 1.86 greater, so it seems to confirm the assumption of rate dependence on the number of jets.
For hard $\gamma^{B/W}\gamma^{B/W}$ scattering, the rate of $(2.13 \pm 0.28)\cdot10^{-5}$ was obtained.
All uncertainties are statistical, calculated as $\sigma_i = \sqrt{\epsilon_i (1-\epsilon_i)/N_i}$, where $\epsilon_i$ is the respective vertex finding rate, and $N_i$ is the total number of events in a sample.

\begin{figure}[bt]
	    \centering
	 	 \begin{subfigure}{0.49\textwidth}
	 	 	\centering
	 	 	\includegraphics[width=\figwidth]{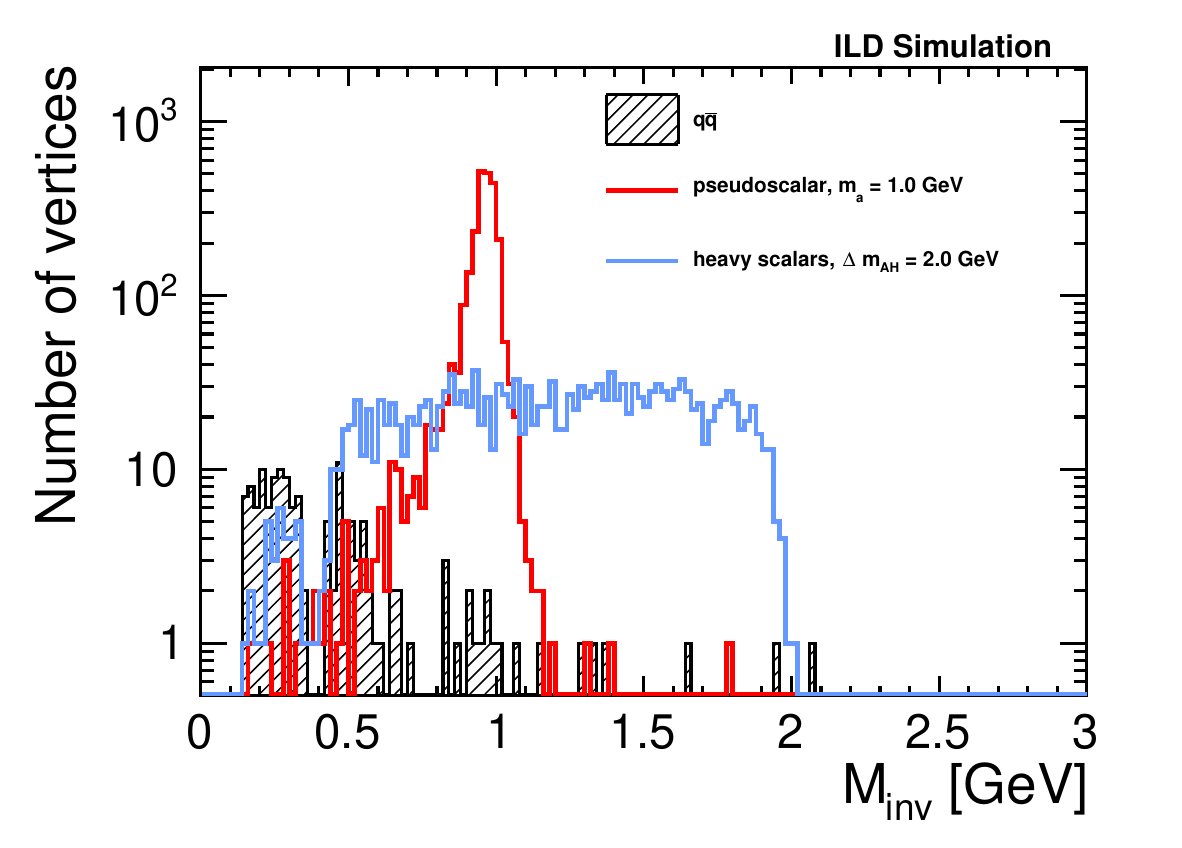}
	 	 \end{subfigure}%
	 	 \begin{subfigure}{0.49\textwidth}
	 	 	\centering
	 	 	\includegraphics[width=\figwidth]{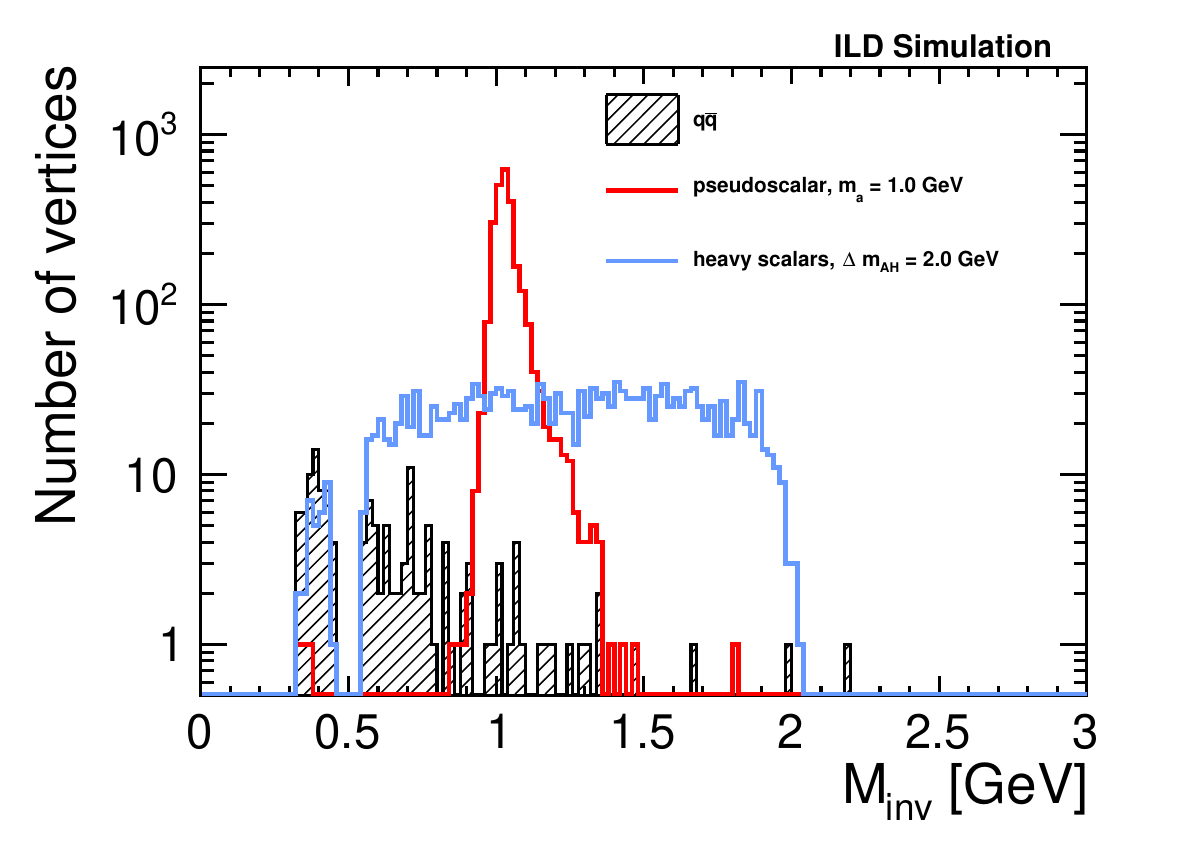}
	 	 \end{subfigure}
	  \vspace*{\figskip}
 	 \caption{Distributions of the invariant mass of tracks coming out of a displaced vertex, assuming tracks are electrons (left) and pions (right). Black histograms correspond to the q$\bar{\mathrm{q}}$ sample and different colors to various signal scenarios -- pseduoscalar with $m_a=1$\,GeV (red) and scalars with $\Delta m_{AH}=2$\,GeV (azure). All histograms are normalized to the number of simulated MC events. }
	 	 \label{fig:qq_vs_sig_mass}
\end{figure}

\begin{figure}[bt]
	    \centering
	 	 \begin{subfigure}{0.49\textwidth}
	 	 	\centering
	 	 	\includegraphics[width=\figwidth]{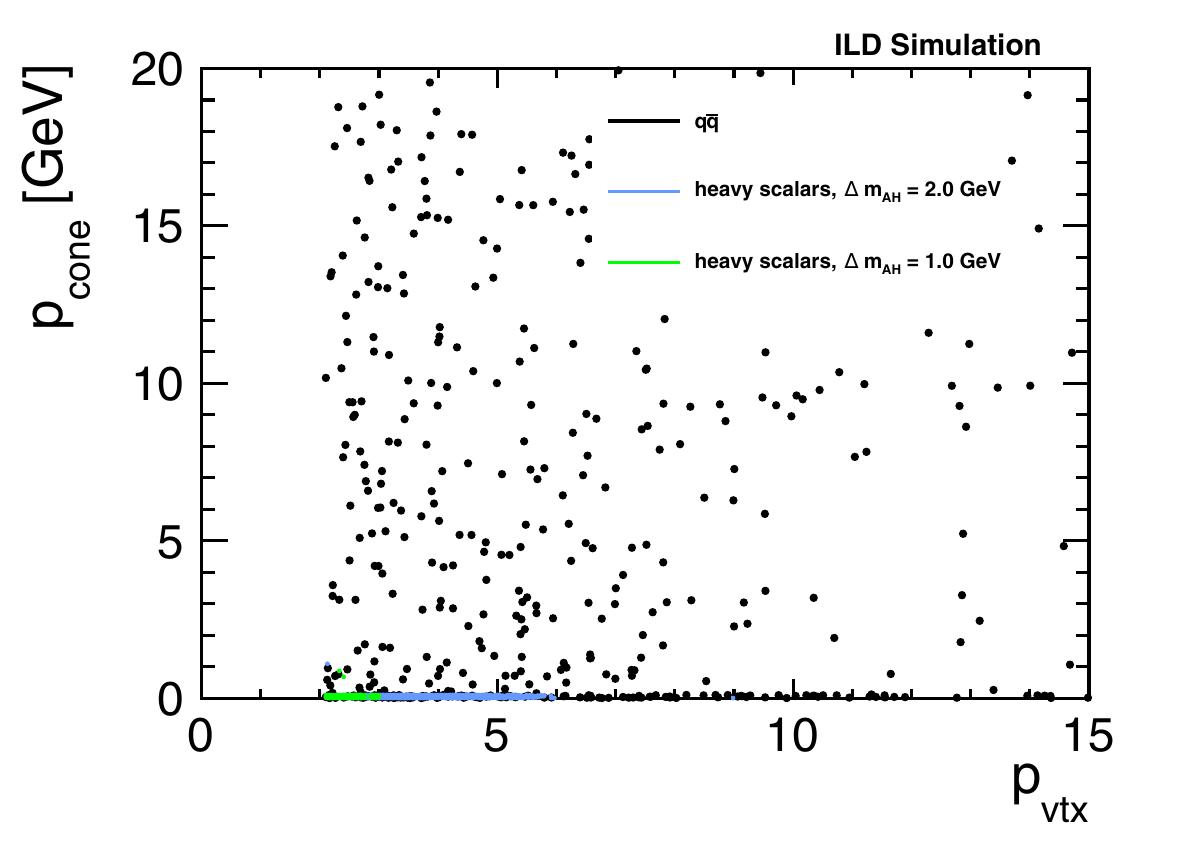}
	 	 \end{subfigure}%
	 	 \begin{subfigure}{0.49\textwidth}
	 	 	\centering
	 	 	\includegraphics[width=\figwidth]{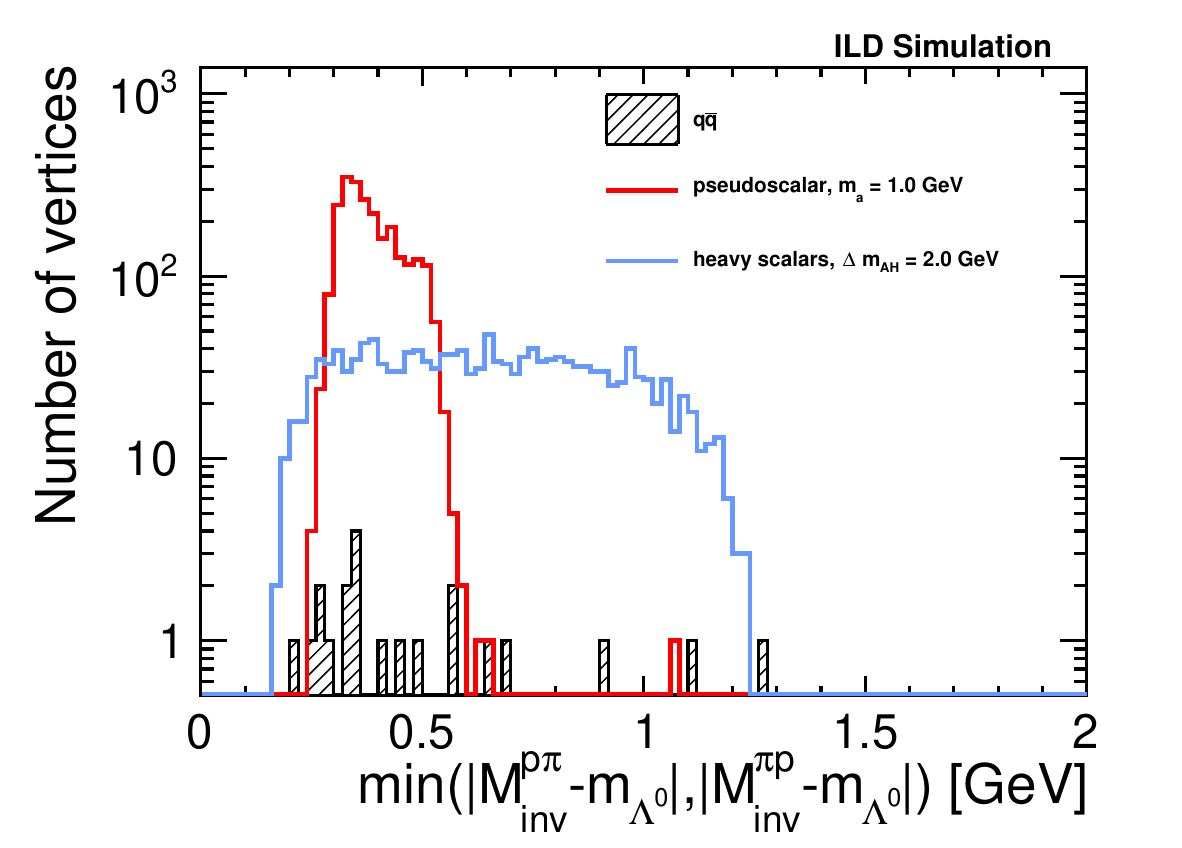}
	 	 \end{subfigure}
	  \vspace*{\figskip}
 	 \caption{Left: momentum $p_{vtx}$ of the track pair system at a displaced vertex shown as a function of the sum of magnitudes of track momenta, $p_{cone}$, in a cone surrounding the total momentum of track pair coming out of the vertex. Black dots correspond to the \qqbar background, green to scalars with $\Delta m_{AH}=1$\,GeV and azure to scalars with $\Delta m_{AH}=2$\,GeV. Right: distribution of the difference between $m_{\Lambda^0}$ and the invariant mass of a track pair assuming that one of them is a pion and the other is a proton, where the smaller of two possible values is taken. The black histogram correspond to the \qqbar background, red to pseduoscalar with $m_a=1$\,GeV and azure to scalars with $\Delta m_{AH}=2$\,GeV. All histograms are normalized to the number of simulated MC events. }
	 	 \label{fig:qq_vs_sig_other}
\end{figure}

Fully hadronic or leptonic decays of V$^0$ particles are efficiently identified and rejected by the cuts described so far. However, this is not the case for semileptonic K$^0_{\mathrm{L}}$ decays, since they involve neutrinos in the final state, resulting in smeared and shifted mass distributions. A similar effect can be due to large reconstruction uncertainties, for short or collimated tracks in particular. This is visible on distributions presented in Figure~\ref{fig:qq_vs_sig_mass} for hypotheses that the tracks are electrons (left) and pions (right) after the standard selection.  
To further improve background rejection and final sensitivity to LLPs, we also define a \textit{tight} selection with additional cuts. In this case, cuts in the electron- and pion-track mass hypotheses are extended to exclude all candidates with $M_{inv}^{ee},M_{inv}^{\pi\pi}<700$\,MeV. 
The window in the cut on $\Lambda^{0}$ mass is reduced from $\pm50$\,MeV to $\pm20$\,MeV, since now it can significantly improve the signal selection efficiency without much increase in the background. 
In the tight selection, we also use the fact that final state tracks in the signal processes should be quite isolated from other activity in the event, while for the background, displaced vertices are mostly reconstructed inside the jets. The isolation measure is chosen as a scalar sum $p_{cone}=\Sigma_i |\vec{p}_i|$ of the track momenta in a cone spanned by an opening angle $\beta$ of $\cos{\beta}>0.98$, around the combined momentum $p_{vtx}$ of the track pair at the vertex. 
Tracks whose first or last hit is within 30\,mm of the vertex candidate are not included in the sum to avoid excluding vertices with more than two tracks in the final state.
The distribution of $p_{cone}$ as a function of $p_{vtx}$ is presented in Figure~\ref{fig:qq_vs_sig_other} (left) for the \qqbar background and $\Delta m_{AH}=2,3$\,GeV signal scenarios after the standard selection.
To suppress background from displaced vertices reconstructed inside hadronic jets we require that $p_{cone}<1$\,GeV. This is a conservative requirement, since $p_{cone}=0$ in the vast majority of signal events.

Because of the isolation criterion, the tight selection might be seen as partly model-dependent (some models could predict an LLP production within or close to hadronic jets). For this reason it is considered separately from the standard set of cuts. However, it is important to note that the vast majority of models currently considered do not predict an LLP production in association with close-by high-momentum tracks. 
Tight cuts on the $M_{inv}^{ee},M_{inv}^{\pi\pi}$, and on the isolation, make the cut on $\Lambda^0$ mass redundant. This is visible in Figure~\ref{fig:qq_vs_sig_other} (right), where  the mass difference distribution after the tight selection, but without the cut on the $\Lambda^0$ mass window, is shown for hypotheses that one track is a pion and the other is a proton. 

As mentioned in Section~\ref{sec:framework}, the SM background samples had to be reprocessed due to the modified reconstruction criteria. For this reason, \qqbar events visible in Figures~\ref{fig:qq_vs_sig_mass} and \ref{fig:qq_vs_sig_other} correspond only to the luminosity of 1.75\,fb$^{-1}$ and the histograms have not been rescaled for better readability. However, only the shapes of the distributions should be compared in these plots and not the relative numbers of vertices, since we do not assign any particular cross sections to the analyzed signal scenarios. The total number of background events expected in the actual experiment is taken into account later.

The tight selection reduces the displaced vertex finding rate by over an order of magnitude;
see Tab.~\ref{tab:sm_bg} for selection rate comparison. 
We also find that the tight selection does not provide any improvement in the removal of overlay events.
The price for an improvement in background rejection is a loss of sensitivity to the most challenging signal scenarios, with $\Delta m_{AH}=1$\,GeV and, in particular, $m_a = 300$\,MeV, which suffer from the cuts on the invariant mass.

\begin{table}[bt]
    \centering
    \caption{The vertex finding rates directly obtained for the high-$p_T$ SM backgrounds inside the TPC region after different sets of cuts. Large statistical uncertainties for tight selection rates result from small number of MC events remaining after the selection. The total number, $N_{bg}$, of background events expected after both selections, taking into account all channels presented in Table~\ref{tab:cross_sec}, is also shown in the last column. $N_{bg}$ is calculated assuming the total luminosity of 2\,ab$^{-1}$ (for details, see Sec.~\ref{sec:limits}). }
    \label{tab:sm_bg}
    \setlength{\tabcolsep}{4pt}
    \renewcommand{\arraystretch}{1.3}
    \begin{tabular}{cccc|c} \hline
          background channel&  qq&  qqqq& $\gamma^{B/W}\gamma^{B/W}$& $N_{bg}$\\ \hline
        Finding rate (standard) &  $(7.99\pm0.68)\cdot10^{-4}$&  $(1.486\pm0.094)\cdot10^{-3}$&  $(2.13 \pm 0.28)\cdot10^{-5}$& 128,804\\ 
         Finding rate (tight) &  $(2.30\pm1.15)\cdot10^{-5}$&  $(3.57\pm1.46)\cdot10^{-5}$& $(1.06\pm0.61)\cdot10^{-6}$& 3,763\\ \hline
    \end{tabular} 
\end{table}

\section{Vertex finding results \label{sec:vtx_finding}}

\begin{figure}[bt]
	    \centering
	 	 \begin{subfigure}{0.49\textwidth}
	 	 	\centering
	 	 	\includegraphics[width=\figwidth]{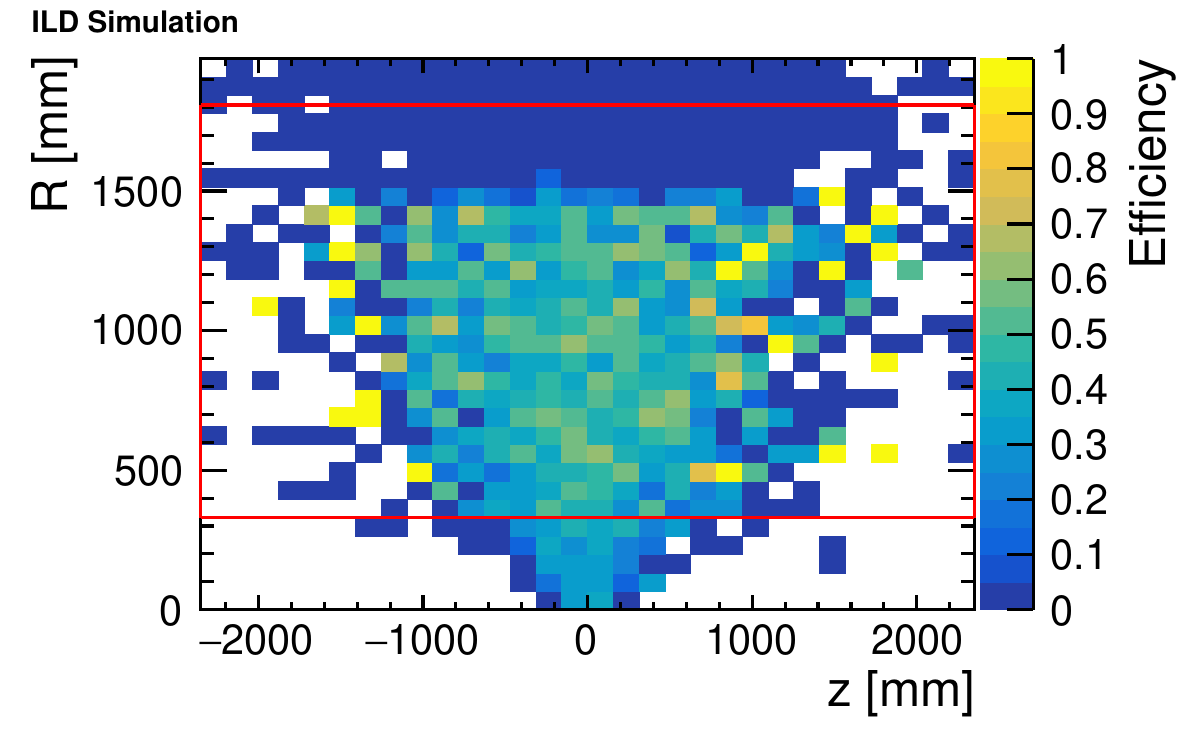}
	 	 \end{subfigure}%
	 	 \begin{subfigure}{0.49\textwidth}
	 	 	\centering
	 	 	\includegraphics[width=\figwidth]{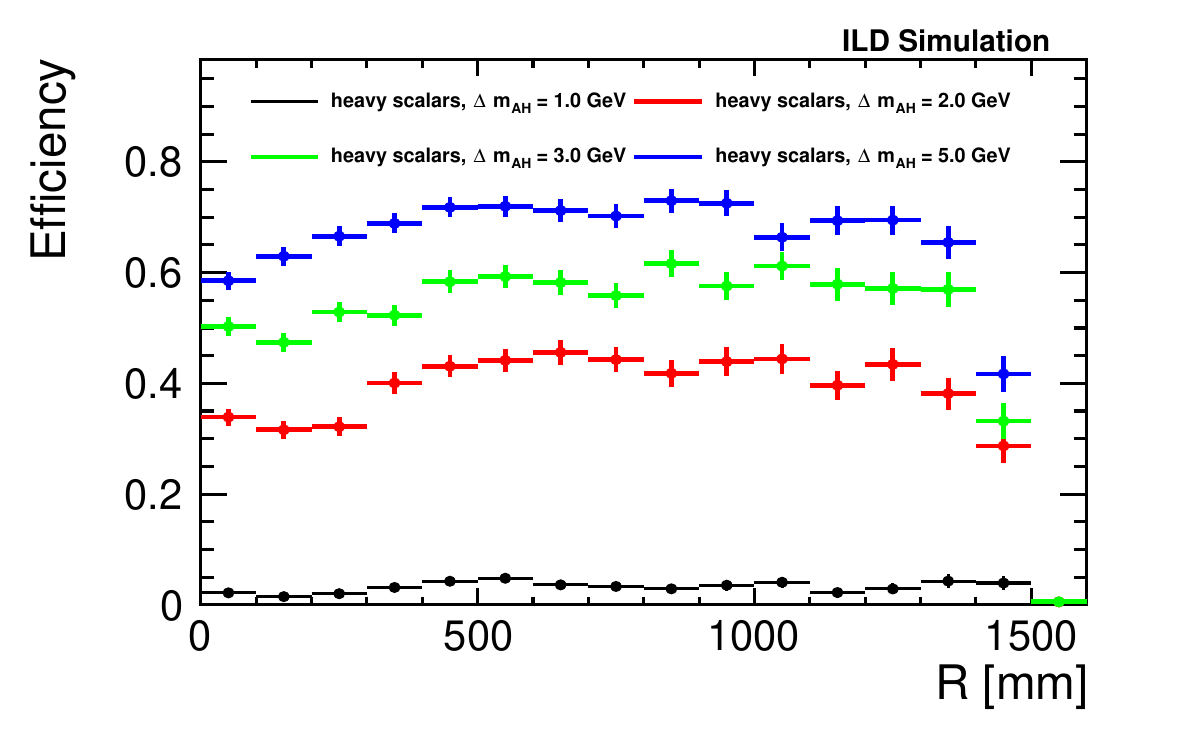}
	 	 \end{subfigure}
	  \vspace*{\figskip}
 	 \caption{Vertex finding efficiency in the scalar pair-production scenario after the standard selection, but without the cut on the vertex radius. Left: the efficiency as a function of the true LLP decay vertex position in the detector for $\Delta m_{AH}=2$\,GeV; the TPC volume is shown with the red box. Right: the efficiency as a function of the true LLP decay vertex radius for all scenarios considered in the analysis.}
	 	 \label{fig:res_idm}
\end{figure}

The resulting vertex finding efficiency after selection described in the previous sections, as a function of the true LLP decay vertex radius and $z$ coordinate, is shown in Figure~\ref{fig:res_idm} (left) for the $\Delta m_{AH}=2$\,GeV heavy scalar scenario. The vertex is considered ``correct'' if it is reconstructed within 30\,mm of the true vertex. In Figure~\ref{fig:res_idm} (right) the efficiency is presented as a function of the true decay vertex radius for all the heavy scalar scenarios considered. For the plots presented, cuts on the vertex position were not applied, to indicate differences between the efficiency for LLP decays inside the TPC region and the silicon tracker ($R<329$\,mm). It is noticeable that for soft final states the vertex finding efficiency is higher within the TPC volume than in the silicon tracker, which is a direct consequence of a higher track reconstruction efficiency for highly displaced soft tracks~\cite{Klamka:2023kmi}.

\begin{figure}[bt]
	    \centering
	 	 \begin{subfigure}{0.49\textwidth}
	 	 	\centering
	 	 	\includegraphics[width=\figwidth]{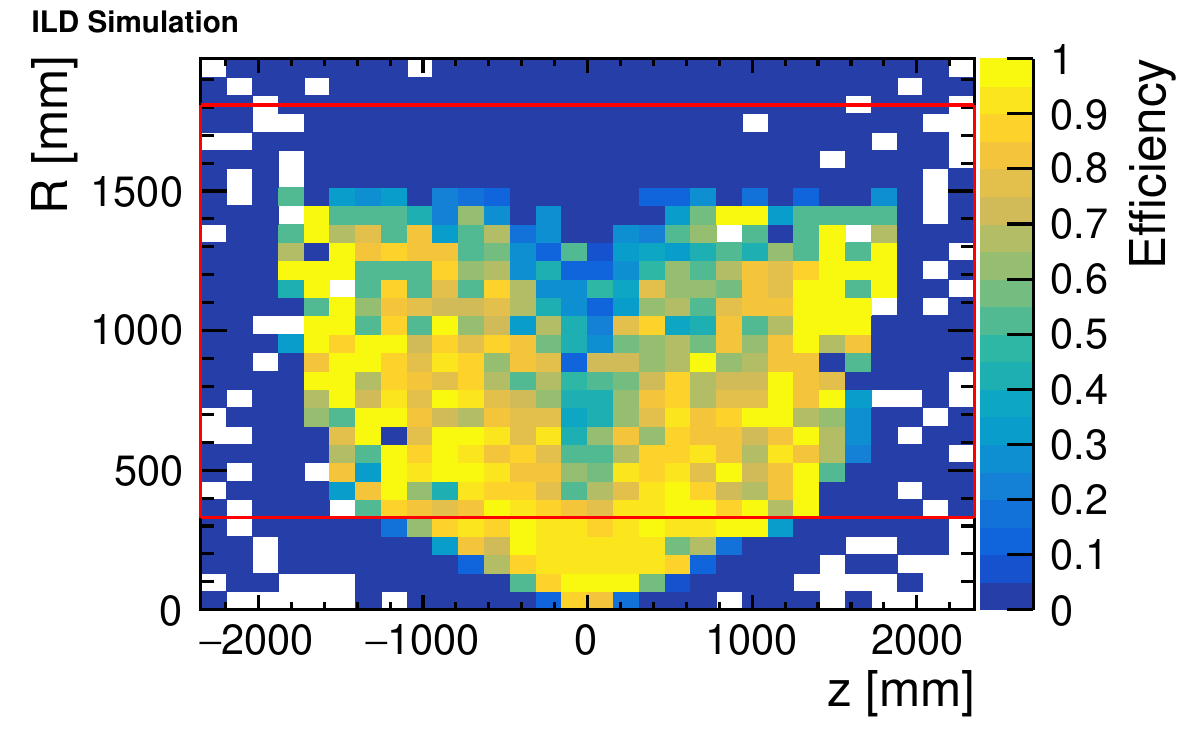}
	 	 \end{subfigure}%
	 	 \begin{subfigure}{0.49\textwidth}
	 	 	\centering
	 	 	\includegraphics[width=\figwidth]{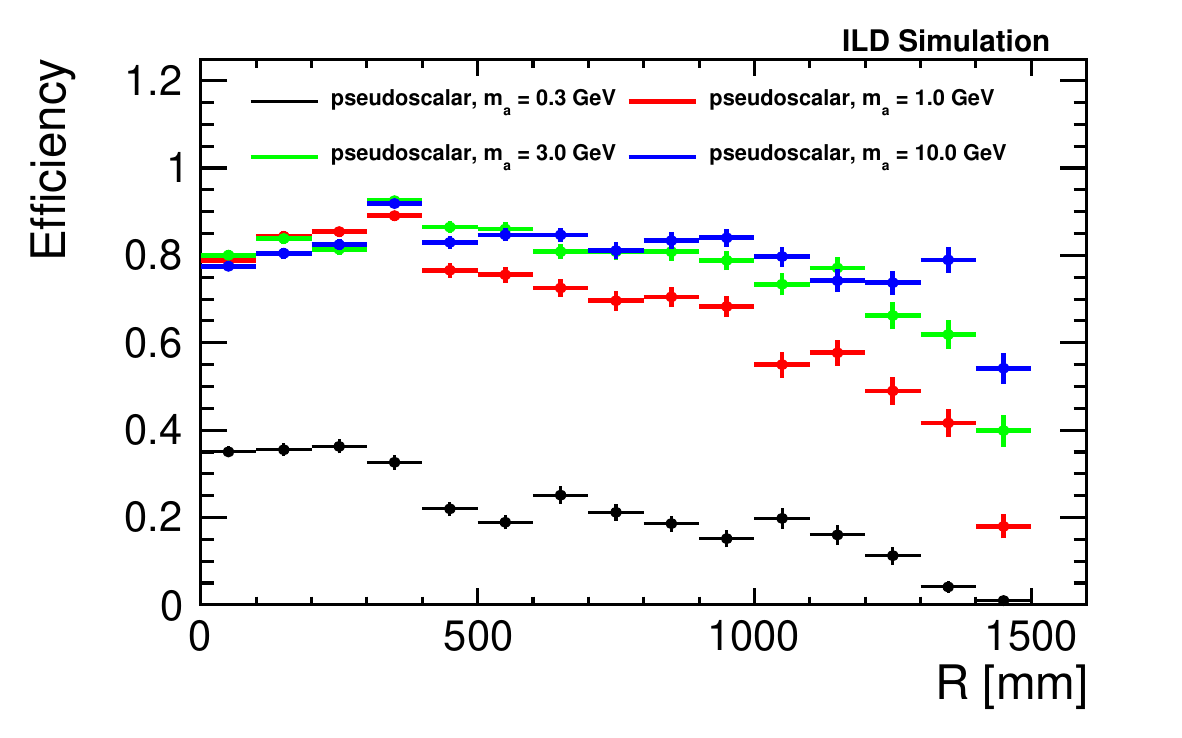}
	 	 \end{subfigure}
 	 \caption{Vertex finding efficiency in the light pseudoscalar production scenario after the standard selection, but without the cut on the vertex radius. Left: the efficiency as a function of the true LLP decay vertex position in the detector for $m_a = 1$\,GeV; the TPC volume is shown with the red box. Right: the efficiency as a function of the true LLP decay vertex radius for all scenarios considered in the analysis.}
	 	 \label{fig:res_alp}
\end{figure}

The corresponding vertex finding results for light pseudoscalar production are presented in Figure~\ref{fig:res_alp}, for the $m_a = 1$\,GeV scenario (left), and compared for all the benchmarks considered (right). Here, in contrary to the heavy scalar case, for the low-mass scenarios the efficiency is higher in the region of silicon detectors (vertex detector, VTX, and silicon internal tracker, SIT), closer to the beam pipe. This is due to very high collinearity of tracks from the LLP decay; in the TPC, hits close to the decay vertex merge, while in a silicon tracker higher resolution allows for a better track separation and more accurate vertex reconstruction.
However, it should be noted, that the TPC still greatly enhances the detector acceptance, since in an all-silicon tracker the reconstruction is limited by a number of layers, regardless of the final state kinematics.
The dip in efficiency visible in the central part of the detector is an effect of the cuts on  $\phi_{1} / \phi_{arc}$ for $z_{12} < 100$\,mm. If hits close to the vertex merge, one of the tracks has a starting point somewhere along the second track, which looks like those two tracks are intersecting, rather than coming out of a single point.

The final vertex finding efficiencies inside the TPC, after both standard and tight sets of cuts, for the considered scalar and pseudoscalar production scenarios, are presented in Table~\ref{tab:results}. For the scalars, the efficiency strongly depends on their mass splitting (Z$^*$ virtuality), which is responsible for the final state boost. Good performance is achieved for $\Delta m_{AH}\geq2$\,GeV, while for the $\Delta m_{AH}=1$\,GeV scenario it suffers mostly from the $p_T^{vtx}>1.9$\,GeV cut used to suppress the overlay background. For the light pseudoscalars the efficiency dependence is opposite, i.e. it decreases with the final state boost. 
This is due to the high collinearity of tracks coming from decays of very light particles, the reconstruction of which is more efficient in the silicon detectors (VTX and SIT) because of their better two-hit separation.
High performance was obtained for $m_a=1$\,GeV and larger masses; the results for the $m_a=300$\,MeV benchmark are limited by the cuts used to suppress background from semileptonic K$^0$ decays and misreconstructed photon conversions. It is important to note that the sensitivity in both $\Delta m_{AH}=1$\,GeV and $m_a=300$\,MeV scenarios could be significantly enhanced by a dedicated, model-dependent approach, using additional information about the missing energy or the hard photon produced in association with an ALP.

\begin{table}[bt]
    \centering
    \caption{The vertex finding efficiency inside the TPC region obtained in the analysis after different sets of cuts, both for scalars pair-production and light pseudoscalars for all considered scenarios.}
    \label{tab:results}
    \begin{tabular}{ccccc} \hline
          $\Delta m_{AH}$ [GeV]&  1&  2&  3& 5\\ \hline
         Efficiency (standard) [\%]&  3&  33.2&  43.4& 51.1\\ 
         Efficiency (tight) [\%]&  0.4&  28.3&  40.7& 50.2\\ \hline
         $m_{a}$ [GeV]&  0.3&  1&  3& 10\\ \hline
         Efficiency (standard) [\%]&  7.4&  48.4&  61.7& 65.8\\ 
         Efficiency (tight) [\%]& --&  47.3&  61.7& 65.8\\ \hline
    \end{tabular} 
\end{table}

\section{Limits on the LLP production cross section \label{sec:limits}}

Based on the obtained signal reconstruction efficiencies and expected background levels, the 95\% C.L. limits on the signal production cross section are calculated. We assumed the total integrated luminosity of 2\,ab$^{-1}$ collected in 10 years, 
with a relative share per beam helicity as specified in the ILC H-20 scenario~\cite{Bambade:2019fyw}. 
This corresponds to $9.7\cdot10^{11}$ BXs collected in the experiment, assuming running with nominal  luminosity.  The 1.55 hadronic overlay events per BX on average, and incoherent \epem pairs overlaid on each BX (for details see Sec. \ref{sec:overlay}), result in a total of approximately $2.47\cdot10^{12}$ overlay events. 
We also use the luminosity for the hard $\gamma^{B}\gamma^{B}$ ($\gamma^{B}\gamma^{W},\gamma^{W}\gamma^{B}$) interactions of $43\%$ (53\%) of the total integrated \epem luminosity~\cite{berggren_private_communication}.
The final number $N_{bg}$ of overlay and hard SM background events remaining after standard selection is 312 and 128,804, respectively, as calculated using the reduction factors obtained in Sections~\ref{sec:overlay} and \ref{sec:sm_bg}.

To estimate limits for different lifetimes without generating and processing new samples, signal event re-weighting is performed in order to estimate the limits for a set of LLP lifetimes for each benchmark scenario.  For any given mean lifetime $\tau'$, an event with LLP decay after a time $t_i$ is re-weighted by a ratio $w_i=P(t_i,\tau')/P(t_i,\tau_0)$ of probabilities $P$ from the exponential distribution, where the mean lifetime $\tau_0$ was used to generate an event sample.  
With the number $N_{p}$ of events containing vertices that match to the MC vertex and pass through the selection, 
a fraction $\epsilon_{sel}=N_{p}/N_{\text{total}}$ must be calculated to obtain the limit. However, for very long $\tau'$, \mbox{$c\tau' \gg $ 1\,m,} the number of decays with a large decay length $ct_i \sim {\cal O}(c\tau')$  in a generated sample can be much smaller than the expected number of these events, leading to large event weights and large statistical uncertainties. Therefore, we define a cutoff distance $L_{max}=3$\,m, above which finding a vertex is impossible. Then, with $w_{max}$ being the total probability that an LLP decays before $L_{max}$ (calculated from the exponential distribution for a given $\tau'$), $N_{\text{total}}=\Sigma_i w_i / w_{max}$, where only events with LLP decays before $L_{max}$ are considered, and $N_p=\Sigma_i w_i$, where selection cuts are also applied.

\begin{figure}[bt]
	    \centering
	 	 \begin{subfigure}{0.49\textwidth}
	 	 	\centering
	 	 	\includegraphics[width=\figwidth]{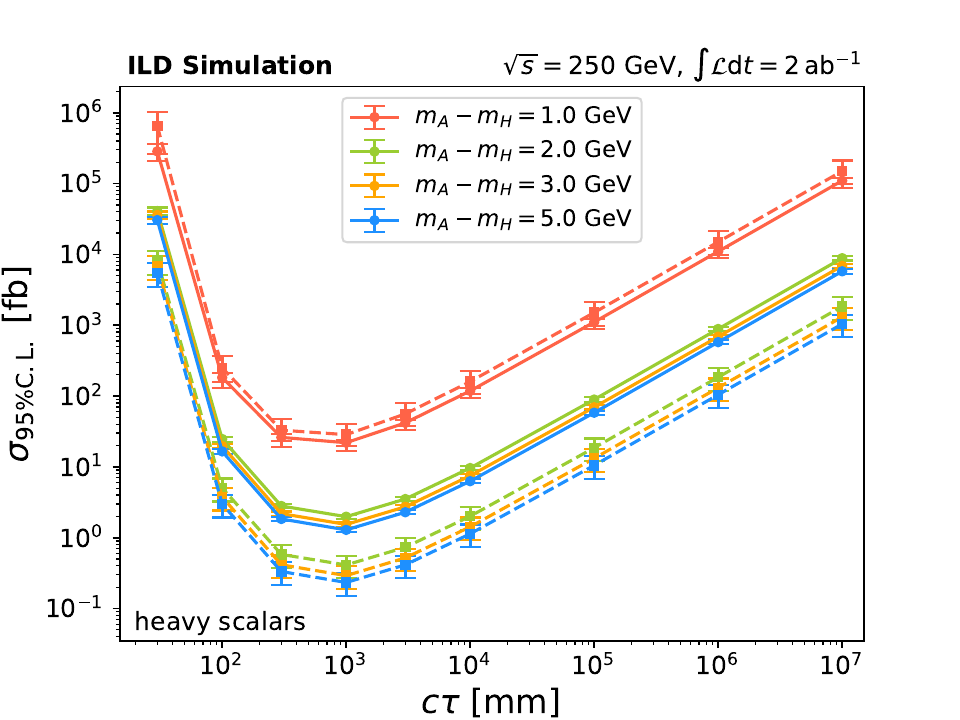}
	 	 \end{subfigure}%
	 	 \begin{subfigure}{0.49\textwidth}
	 	 	\centering
	 	 	\includegraphics[width=\figwidth]{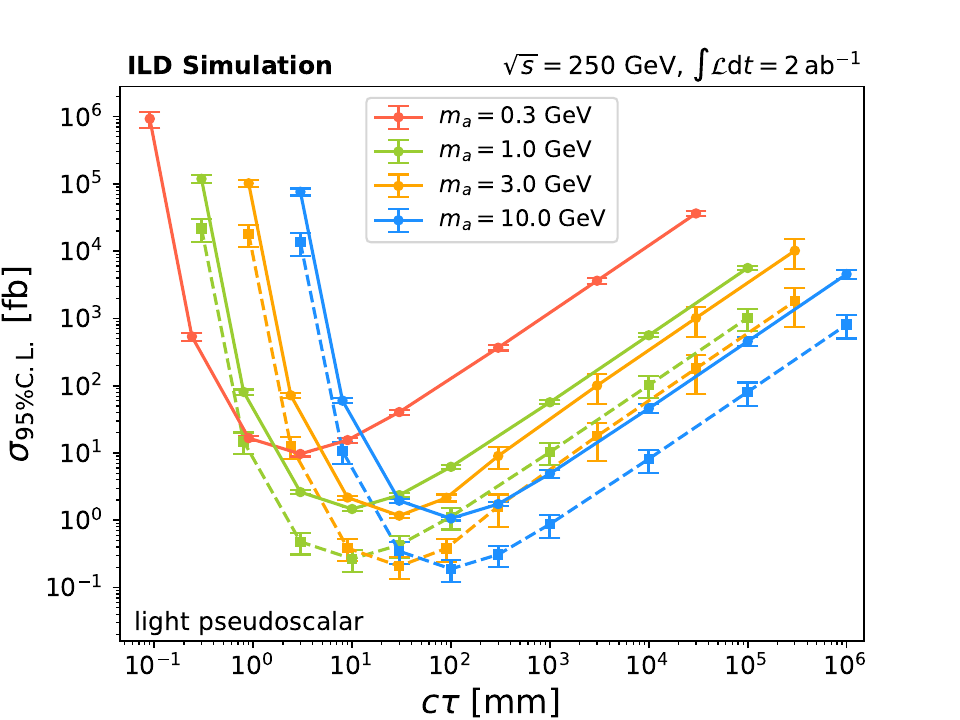}
	 	 \end{subfigure}
  \vspace*{\figskip}
	 	 \caption{Expected 95\% C.L. upper limits on the signal production cross section for the considered benchmarks and different LLP mean decay lengths, for the scalar pair-production (left) and the light pseudoscalar production (right) at $\sqrt{s}=250$\,GeV. Solid lines corresponds to the standard selection and dashed lines to the tight set of cuts. The uncertainties are statistical. }
	 	 \label{fig:limits}
\end{figure}

The 95\% C.L. limit $\sigma_{95\% \mathrm{C.L.}}$ corresponds to the number $N_{lim}=1.96\sqrt{N_{bg}}/\epsilon_{sel}$, where the factor 1.96 corresponds to the unified frequentist interval~\cite{Feldman:1997qc} for the measured number of events equal to the nominal expectation (which is equivalent to the CLs method), assuming that $\epsilon_{sel}N_{lim} \ll N_{bg}$ and a Gaussian distribution for the background uncertainty.
The limits for each of the considered benchmarks in the scalar pair-production (left) and the light pseudoscalar production (right) scenarios are presented in Figure~\ref{fig:limits}. A range of mean decay lengths $c\tau$ is considered and two sets of limits are presented -- for the standard and the tight selection. Also indicated in the figure are the statistical uncertainties resulting from the MC event selection for the signal and all background sources considered. Due to the relatively small size of the reprocessed MC event samples, the dominant contribution comes from estimates of physical event backgrounds, as described in Sec.~\ref{sec:sm_bg}.

In the case of scalar pair-production, good sensitivity can be observed in the range 0.3-10\,m for $\Delta m_{AH}\geq2$\,GeV. This can be extended by the tight selection, which improves the limits by an order of magnitude. For the $\Delta m_{AH}=1$\,GeV scenario, the limits get slightly degraded by the tight selection and go down to the level of tens of femtobarns in the 0.3-3\,m range.

For the light pseudoscalars, limits at the order of femtobarns can be obtained with the standard selection for $m_a\geq1$\,GeV in a 3-1000\,mm range, depending on the scenario. The tight selection improves the limit by up to an order of magnitude, allowing to reach the level of femtobarns (or lower) even up to $c\tau\simeq10$\,m. However, it causes a complete loss of sensitivity to the $m_a=300$\,MeV benchmark, which can be constrained down to $\mathcal{O}$(10\,fb) level using the standard selection. 

\section{Conclusions \label{sec:summary}}

We demonstrate the potential of the ILD experiment to search for neutral LLP production using a displaced vertex signature. The study is based on full detector simulation and conducted from the experimental perspective, with two types of benchmarks. The first involves scalar pair-production, where one of the scalars is long-lived and decays to SM particles and DM. The LLP mass is set to 75\,GeV and the mass splitting between the LLP and DM is chosen to be at the $\mathcal{O}$(1 GeV) level, resulting in soft, non-pointing tracks in the final state. The second benchmark type is the production of a very light pseudoscalar with an $\mathcal{O}$(1 GeV) mass, that decays into highly boosted, almost colinear SM particles. Such a choice of benchmarks, that are not fully examined at the LHC, tests the detector performance and reconstruction techniques in very challenging conditions. For each signature, we select four signal scenarios, which are motivated by experimental and kinematic properties, rather than by existing constraints in the parameter space of a specific model.   

A number of background processes is considered, both from the beam-beam interactions at a linear collider and hard \epem scattering. We find a set of cuts that suppress backgrounds from the beam-beam events and other SM processes, keeping the model-independent approach.  In addition, a tight selection is also proposed to improve the background rejection, slightly reducing the generality of the approach, but improving the sensitivity for most of the signal scenarios considered in the analysis. Finally, based on these selections, we set expected 95\% C.L. limits on the signal production cross section. For the pair production of scalars, high reach is observed in 0.3-10\,m range from mass splittings of 2\,GeV between the LLP and the DM. Good sensitivity can also be reached for pseudoscalar masses from 1\,GeV, for mean decay lengths in 0.003-1\,m range, depending on LLP mass.

The results presented should be considered the most conservative ones.
For specific model scenarios, stronger limits (for smaller masses and mass splittings in particular) could be obtained by exploiting the expected final state topology and kinematic constraints of the model.
This is, however, beyond the scope of this paper. 

This is the first analysis to address the direct detection of neutral LLPs at colliders following such a general approach, with limits derived in a fully model-independent manner. As a consequence, there are no other results suitable for a direct comparison. 
In addition, only a few existing studies on the potential for LLP detection at future colliders are based on full simulation~\cite{Kucharczyk:2022pie,Zhang:2024bld,Jeanty:2022cwr}. These studies focus on exotic Higgs decays at CLIC, CEPC, and the SiD detector for the ILC, but rely on more specific signal signatures or present results within a model-specific framework. While the approach described here could be adapted to study similar scenarios, such extensions are left for future work. On the other hand, comparisons with studies based on fast simulation would also be invalid, as this analysis incorporates a much more detailed description of the background, accounting for all potential sources.

\section*{Acknowledgements}

The work was supported by the National Science Centre (Poland) under OPUS research project no.~2021/43/B/ST2/01778. We would like to thank the LCC generator working group and the ILD software working group for providing the simulation and reconstruction tools and producing the Monte Carlo samples used in this study. The authors thank Tania Robens for discussions on macroscopic lifetimes in the IDM. We also thank Susanne Westhoff and Ruth Sch\"afer for all help with the model involving ALPs. This work has benefited from computing services provided by the ILC Virtual Organization, supported by the national resource providers of the EGI Federation and the Open Science GRID. 

\printbibliography

\end{document}